\documentclass[a4paper,fleqn,usenatbib]{mnras}
\def\paren#1{\left( #1 \right)}

\def\ltsima{$\; \buildrel < \over \sim \;$}
\def\lsim{\lower.5ex\hbox{\ltsima}}
\def\gtsima{$\; \buildrel > \over \sim \;$}
\def\gsim{\lower.5ex\hbox{\gtsima}}
\usepackage{newtxtext,newtxmath}
\usepackage{empheq}
\usepackage[T1]{fontenc}
\usepackage{ae,aecompl}
\usepackage{placeins}
\usepackage{verbatim}
\usepackage[none]{hyphenat}
\usepackage[utf8]{inputenc}
\usepackage{booktabs}

\usepackage{graphicx}	
\usepackage{amsmath}	
\usepackage{amssymb}	
\title[BH Mergers induced by Tidal Encounters]{Black Hole Mergers Induced by Tidal Encounters with a Galactic Centre Black Hole}

\author[Joseph John Fernández, Shiho Kobayashi]{
Joseph John Fernández$^{1}$, Shiho Kobayashi$^{1}$
\\
$^{1}$Astrophysics Research Institute, LJMU, IC2, Liverpool Science Park, 146 Brownlow Hill, Liverpool L3 5RF, UK}

\date{Accepted XXX. Received YYY; in original form ZZZ}

\pubyear{2018}

\begin{document}
\label{firstpage}
\pagerange{\pageref{firstpage}--\pageref{lastpage}}
\maketitle
\begin{abstract}
We discuss the properties of stellar mass black hole (BH) mergers induced by tidal encounters with a massive BH at galactic centres or potentially in dense star clusters. The tidal disruption of stellar binaries by a massive BH is known to produce hypervelocity stars. However, such a tidal encounter does not always lead to the break-up of binaries. Since surviving binaries tend to become hard and eccentric, this process can produce BH mergers in principle.  For initially circular binaries, we show that the gravitational wave (GW) merger times become shorter by a factor of more than $10^{2}$ ($10^5$) in $10\%$ ($1\%$) of the surviving cases. This reduction is primarily due to the  growth in binary's eccentricity at the tidal encounter. We also investigate the effective spins of the survivors,  assuming that BH spins are initially aligned with the binary orbital angular momentum. We find that binary orientations can flip in the opposite direction at the tidal encounter. For the survivors with large merger time reduction factors, the effective spin distribution is rather flat. 
We estimate the merger rate due to the tidal encounter channel to be $\sim 0.6\ \textrm{Gpc}^{-3}\textrm{yr}^{-1}$.
This mechanism is unlikely to be the dominant formation channel of BH mergers. However, the current and near-future GW observatories are expected to detect an enormous number of BH mergers.
If mergers are found in the vicinity of massive BHs (e.g. the detection of GW lensing echoes or preceding extreme-mass-ratio bursts), this mechanism would provide a possible explanation for their origin. 
\end{abstract}


\section{Introduction}
The recent LIGO/Virgo observations mark the dawn of the gravitational wave (GW) astronomy. The successive detections of GW signals from black hole (BH) mergers suggest that BH-BH binaries are primary sources for ground-based GW detectors \citep{ligo1, ligo2, lv1, ligo3, det5}. 
Several formation scenarios have been discussed so far to explain their origin, and the scenarios can be roughly classified in two groups: 1) isolated field binary models such as the classical field binary formation model, homogeneous chemical evolution and massive overcontact binaries, e.g. \cite{chris1, mandelmink, marchantApj}, and 2) dynamical formation models such as a sequence of three-body interactions in globular clusters or nuclear star clusters \citep{rasio1, rasio3, rasio2, arcaNHT}, the Kozai-Lidov mechanism \citep{secev, klm, antImpLigo, eccKozaiStephan, mergerskozai}, or binary hardening in AGN disks \citep{mergerinnuc}. 

With further improvements planned for LIGO and Virgo, and other GW detectors (KAGRA, LIGO India) coming online, a large number of BH mergers are expected to be discovered in the coming years. The planned Space GW detectors (e.g. LISA, DECIGO, BBO, MAGIS, ALIA) also should allow us to further study their properties \citep{ligop, virgop, kagrap, indiap, lisap,  decigop, bbop, magisp, aliap}.   It may be possible to identify the signatures of specific formation models in the upcoming sample. 

Tidal disruptions of binaries by a massive BH are well known to produce hypervelocity stars \citep{hills, yu}. However, our previous numerical simulations have revealed that about $10\%$ of binaries can survive even very deep encounters \citep{kobayashi1, brown1}. Most survivors are hard and eccentric, and therefore they have GW merger times much shorter than those of the pre-encounter binaries. As \cite{addison2015} have pointed out, the tidal encounter process could provide a new formation channel of BH mergers in principle. 
In this paper, we investigate the tidal encounter of BH binaries with a massive BH by using the restricted three-body approximation \citep{kobayashi1, brown1}. 
Since the evolution of BH binaries depends only on a small number of parameters  in this approximation, we can provide a clear picture of how the properties of survivors (e.g. the GW merger time, the effective spin) depend on the initial configuration of the system. Although the study in this paper focuses on the tidal encounter
dynamics (the interaction between initially circular binaries and a massive BH), we also discuss the astrophysical implications. 

The structure of the paper is as follows. In section \ref{sec:rtb} we describe the restricted three-body approximation which allows us to efficiently sample the binary parameter space. It is also discussed how the tidal encounter distorts binary orbits. In section \ref{sec:surv} we use Monte Carlo simulations to characterize the distributions of the GW merger times and effective spin parameters of survivors. In section \ref{sec:comp} we briefly discuss the constraints from the current effective spin measurements.  
In section \ref{sec:conc} we give the discussion and conclusions.

\section{Tidal encounter process} \label{sec:rtb} 
\subsection{The restricted three-body approximation}
We consider a BH binary system, of component masses $m_{1}$ and $m_{2}$ ($m=m_{1}+m_{2}$), and 
assume that the centre of mass (COM) approaches a massive BH with mass $M$ 
on a parabolic orbit. 
If the mass ratio is large $M/m \gg 1$, the restricted three-body formalism provides a good 
approximation to evaluate the binary evolution. In this approximation, 
the relative motion of the two binary components $\mathbf{r} \equiv \mathbf{r}_{2}-\mathbf{r}_{1}$ 
is described by the following equation \citep{kobayashi1},
\begin{eqnarray} \label{eq:eqmot}
\frac{d^{2}\mathbf{r}}{dt^{2}} = - \frac{GM}{r_{m}^{3}}\mathbf{r} + 3\frac{GM}{r_{m}^{3}} (\mathbf{r}\cdot \hat{\mathbf{r}}_{m})\hat{\mathbf{r}}_{m} - \frac{Gm}{r^{3}}\mathbf{r},
\end{eqnarray} 
where $\mathbf{r}_1$, $\mathbf{r}_2$ and $\mathbf{r}_{m}$ are the positions of the primary $m_1$, the secondary $m_2$ and the binary's COM relative to the massive BH (i.e. the massive BH is at the origin), and $\hat{\mathbf{r}}_{m}=\mathbf{r}_m/r_m$ is a unit vector. Using the distance $r_p$ of closest approach
(periastron) and the angle $f$ from the point of closest approach (true anomaly), the parabolic orbit
$\mathbf{r}_{m}$ can be expressed as
\begin{eqnarray} \label{eq:rm}
\mathbf{r}_{m} = \frac{2r_{p}}{1+\cos{}f} (\cos{f} ~\hat{\mathbf{x}} +\sin{f} ~\hat{\mathbf{y}}),
\end{eqnarray}
where $\hat{\mathbf{x}}, \hat{\mathbf{y}}$ and $\hat{\mathbf{z}}$ are the unit vectors of
a Cartesian coordinate system. Since we have assumed that the COM orbit is in x-y plane, the z component is zero and omitted in eq (\ref{eq:rm}). The coordinate system is chosen so that $\hat{\mathbf{x}}$ points from the massive BH in the direction of the periastron of the COM orbit.

\begin{figure} 
\centering
\includegraphics[width=0.8\columnwidth]{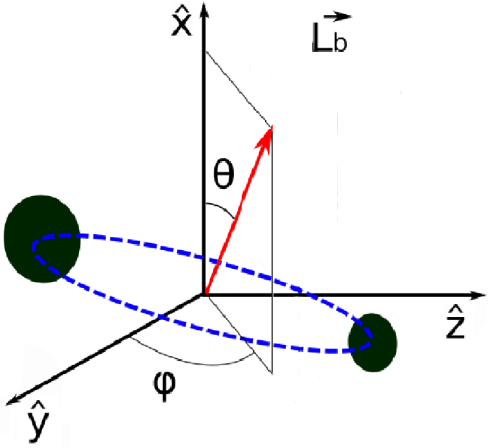}
\caption{  
The binary angular momentum $\hat{\mathbf{L}}_b$
is defined in the rest frame of the binary's COM. The same set of 
unit vectors $\hat{\mathbf{x}}$, $\hat{\mathbf{y}}$ and $\hat{\mathbf{z}}$
is used to represent the axes of a Cartesian coordinate system 
(the coordinate axes in the binary's COM rest frame are 
parallel to those in the massive BH rest frame).
The polar angle $\theta$ is defined as the angle between  $\mathbf{L}_b$ and $\hat{\mathbf{x}}$.  With this parameterisation, the outcome of the tidal encounter does not depend on the azimuthal angle $\varphi$ for $D\ll1$, because the COM moves along the x axis 
in the massive BH rest frame (the parabolic orbit becomes radial for 
$D\ll 1$).}
\label{fig:angles}
\end{figure}

The tidal force of the massive BH overcomes the self-gravity of the binary at the tidal radius $r_{t}=(M/m)^{1/3}a_0$, where 
$a_0$ is the initial binary separation. We define the penetration factor $D=r_{p}/r_{t}$ as a measure of how deeply the binary penetrates into the tidal sphere as it moves along the parabolic trajectory
\footnote{We will consider only the full loss cone case in this paper (see section 3.2). In the empty loss cone case, the picture can be very different, as energy exchange between the inner and outer binary is inefficient but angular momentum can still be exchanged over multiple periapsis passages, leading to eccentricity enhancement and still allowing for enhanced mergers (e.g. \cite{bradnick}).}
We will consider only the full loss cone case in this paper (see section 3.2). In the empty loss cone case, the picture can be very different, as energy exchange between the inner and outer binary is inefficient but angular momentum can still be exchanged over multiple periapsis passages, leading to eccentricity enhancement and still allowing for enhanced mergers (e.g. \cite{bradnick}).. 
Although we need to specify the initial distance of the binary's COM to  the massive BH $r_{m,0}$ to carry out numerical simulations, the binary evolution is largely independent of it if simulations start at a large enough radius $r_{m,0}\gg r_t$. In our run, we assume $r_0=10r_t$ which is sufficient for convergence. The initial binary phase (at $t_0 = t(r_0)<0$) 
$\phi_0=\omega t_0 +\phi$ 
is characterized by using the effective phase $\phi$ at $t=0$ (i.e. 
at the periastron passage), where $\omega$ is the constant angular 
velocity of the binary at $r_m \gg r_t$. Naturally, the actual phase 
at $t=0$ is in general different from $\phi$ due to the tidal force 
of the massive BH. 
If the binary angular momentum $\mathbf{L}_b$ is in the 
z direction (a planar prograde case), the initial binary phase 
$\phi_0$ is the angle between $\mathbf{r}$ and 
$\hat{\mathbf{y}}$ at $t=t_0$. In a general case, we first define
the initial separation vector $\mathbf{r}$ and the initial 
velocity $d \mathbf{r}/dt$ assuming the planar prograde case, and we rotate them as $\mathbf{L}_b$ points to the ($\theta$, $\varphi$) direction (see figure \ref{fig:angles}) before we start to evaluate the temporal evolution of $\mathbf{r}$.

In the restricted three-body approximation, results can be simply rescaled in terms of binary masses, their initial separation, and the binary-to-MBH mass ratio. If the binary is initially circular,
the system is essentially characterized by four parameters: the penetration factor $D$, the effective binary phase $\phi$
and the orientation $(\theta,\varphi)$.
We carry out numerical simulations by using dimensionless quantities $\tilde{\mathbf{r}} = (M/m)^{1/3} \mathbf{r}/r_{p}$ and $\tilde{t}=\sqrt{GM/r_{p}^{3}} t$. With these the equation of motion can be rewritten as 
\begin{eqnarray} \label{eq:eqmot2}
\frac{d^{2}\tilde{\mathbf{r}}}{d\tilde{t}^{2}} = \left(\frac{r_{p}}{r_{m}}\right)^{3} \left[-\tilde{\mathbf{r}} + 3(\tilde{\mathbf{r}}\cdot \hat{\mathbf{r}}_{m}) \hat{\mathbf{r}}_{m}\right] - \frac{\tilde{\mathbf{r}}}{\tilde{r}^{3}}
\end{eqnarray}
To close the system, the temporal evolution of the true anomaly is needed. Using the dimensionless time, this is given by 
\begin{eqnarray} \label{eq:eqmot3}
\frac{df}{d\tilde{t}}= \frac{\sqrt{2}}{4}\left( 1 + \cos{f} \right)^{2}.
\end{eqnarray}

Since BHs are very compact objects, the probability of collisions among binary members and tidal deformations is negligible. The point particle treatment should be adequate. 
Our Newtonian formulation breaks down if the periastron is close to the event horizon scale $r_g$ of the central massive BH or equivalently if $D\lsim (m/M)^{1/3}r_g/a 
\sim 2\times 10^{-3} (a/1\mbox{au})^{-1} (m/60M_\odot)^{1/3}(M/4\times 10^6 M_\odot)^{2/3}$.
Besides for deep encounters, relativistic corrections become important if the initial binary separation is close to the event horizon scales of the binary members. However, in this case, binaries have short GW merger times even before the tidal encounter, and binary hardening processes are not required to produce BH mergers. 

\subsection{Binary deformation due to the tidal encounter}
Previous studies \citep{kobayashi1, brown1} have shown that around $10\%$ of binaries survive very deep encounters, $D\ll 1$, and the survivors tend to become hard and eccentric.  The GW merger time is very sensitive to the binary semi-major axis $a$ and eccentricity $e$, and it is given by  \citep{peters1964}
\begin{eqnarray} \label{eq:mergtime}
\tau_{gw}& \sim& \frac{5}{256}\left( \frac{c^{5} a^{4}}{G^{3}m_{1}m_{2}m}\right) (1-e^{2})^{7/2} \nonumber \\
&\sim& 6.0 \times 10^{3}\left( \frac{m}{60M_{\odot}}\right)^{-3} \left( \frac{a}{1 \mbox{au}}\right)^{4} (1-e^{2})^{7/2} \textrm{Gyrs}
\end{eqnarray}
where an equal mass binary was assumed in the second line and $M_{\odot}$ is the solar mass. For example, a circular binary composed of two $30 M_{\odot}$ BHs initially separated by $a_0=1 \, \textrm{au}$ would not merge within the age of the universe due to GW emission alone. However, the tidal encounter can make the merger time much shorter.  

\begin{figure} 
\includegraphics[width=\columnwidth]{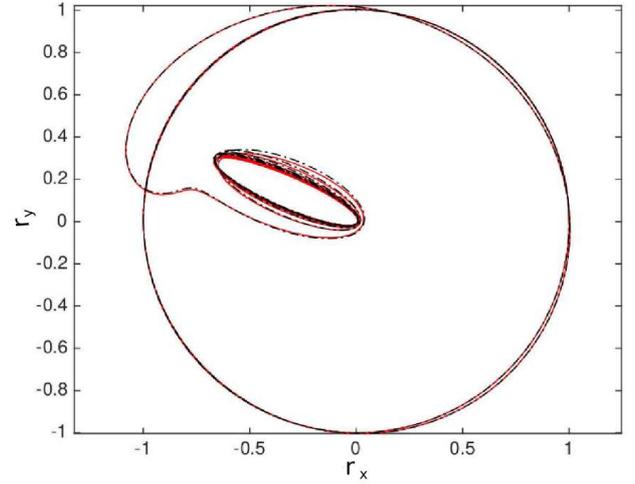}
\caption{The evolution  of binary separation vector $\mathbf{r}=\mathbf{r}_2-\mathbf{r}_1$.
A prograde binary orbit with $D=1$ is assumed to evaluate the restricted three-body approximation orbit (red solid line). The black dashed-dotted line indicate the full three-body orbit. The binary mass ratios are assumed to be $m_{1}/m_{2}=3$ and $M/m = 10^{5}$ for the full three-body calculations. 
Lengths are in units of the initial binary separation $a_0$.}
\label{fig:binexam}
\end{figure}

Figure \ref{fig:binexam} shows an example of a survivor (the red solid line). This is obtained assuming $D=1$ and a prograde orbit (i.e. the angular momentum vectors of the binary components around the binary COM are aligned with the angular momentum of the binary around the massive BH). The semi-major axis of the survivor is smaller by a factor of $2.7$ than that of the initial circular binary, and the survivor is highly eccentric, with $e=0.97$. This leads to a reduction of the merger time by a factor of $\sim 10^{6}$. The black dashed-dotted line indicates the full three-body calculations. The two results are almost identical in the figure, illustrating the accuracy of the restricted three-body approximation.

If the semi-major axis becomes smaller at the tidal encounter, part of the self-binding energy of the binary $\Delta E = (Gm_{1}m_{2}/2)(a^{-1}-a_{0}^{-1})$ is transferred to the orbital energy of the binary COM around the massive BH. This should make the orbit of the COM hyperbolic. 
However, the velocity change $\Delta v$ caused by the released energy ($\Delta E\sim 0.16 G m^{2}/a_{0}$ in the case of figure \ref{fig:binexam}) is much smaller than the original COM velocity $v_m$ around the tidal radius,
\begin{eqnarray}
\frac{\Delta v}{v_m} \lsim \frac{Gm/a_0}{GM/r_t} = \paren{\frac{m}{M}}^{2/3}.
\end{eqnarray}
The orbit around the tidal radius is still very close to the initial parabolic orbit. 
Even if the initial COM orbit is not exactly parabolic, our approximation is still
accurate. Assuming an orbit energy of the COM $E_m=\kappa (Gm^2/a_0)$,
we numerically evaluate the full three body evolution of a binary for $D=1, \theta=0.6\pi, \varphi=0.5\pi, \phi \sim 0.4\pi$,
$M=4\times10^6 M_\odot$ and $m_1=m_2=30M_\odot$. The results are compared with the restricted parabolic approximation results 
for the same set of the four parameters $D, \theta, \varphi$ and $\phi$. 
In both calculations, the binary survives the tidal encounter with the massive BH. 
The differences of the semi-major axis, eccentricity and merger time  
are $\Delta a/a \sim 0.4\%,  4\%$ and $2\%$, $\Delta e/e \sim 0.07\%, 0.6\%$ and $1\%$,
and $\Delta \tau_{gw}/\tau_{gw}\sim 3\%, 20\%$ and $20\%$ for $\kappa=1, 10$ and 100, respectively.  Since the merger time is sensitive to $a$ and $e$, the error in the merger time is rather large for $\kappa \gsim 10$.  However,  for our discussion, only the order-of-magnitude estimate of $\tau_{GW}$ is needed (or a few 10\% error in the $\tau_{GW}$ estimate does not affect our conclusions).  Even for $\kappa=100$ (for which the COM velocity at large distances from the massive BH is about one order of magnitude larger than the binary rotation velocity), the restricted parabolic approximation gives reasonable results.

\subsection{Binary Orientation} \label{subsec:prec}
Corresponding to the change in the binary self-energy, the orientation of the binary is also 
expected to change in general if the binary survives the tidal encounter.
The angular momentum of the binary members around the massive BH is given by 
\begin{eqnarray}
\mathbf{L}= m_{1} \mathbf{r}_{1} \times \mathbf{v}_{1} + m_{2} \mathbf{r}_{2} \times \mathbf{v}_{2},
\end{eqnarray}
where the massive BH is at the origin. Using the binary positions relative to the COM 
$\Delta \mathbf{r}_{1,2}= \mathbf{r}_{1,2}-\mathbf{r}_{m}$, we can rewrite the 
angular momentum as the sum of two components 
$\mathbf{L} = \mathbf{L}_{m} + \mathbf{L}_{b}$ where 
\begin{eqnarray} \label{eq:comam}
\mathbf{L}_{m} &=& m \mathbf{r}_{m}\times \frac{d\mathbf{{r}}_{m}}{dt}, \\
\mathbf{L}_{b} &=& m_{1} \Delta\mathbf{r}_{1} \times \frac{d\Delta \mathbf{r}_{1}}{dt} + m_{2} \Delta\mathbf{r}_{2} \times \frac{d\Delta \mathbf{r}_{2}}{dt}  \nonumber \\
&=&\frac{m_{1}m_{2}}{m} \mathbf{r}\times\frac{d \mathbf{r}}{dt}. 
\end{eqnarray}
The COM angular momentum $\mathbf{L}_{m}$ and the binary angular 
momentum $\mathbf{L}_{b}$ can change at the tidal encounter. However, since 
the binary system moves in the central force field, the total vector $\mathbf{L}$ 
should be conserved. 
Using the equation of motion (\ref{eq:eqmot}), the evolution of $\mathbf{L}_{b}$ is given by 
\begin{eqnarray} \label{eq:torque}
\frac{d \mathbf{L}_{b}}{dt} = \frac{3GMm_{1}m_{2}}{m{r_{m}^{3}}} (\mathbf{r}\cdot \hat{\mathbf{r}}_{m}) \mathbf{r} \times \hat{\mathbf{r}}_{m}.
\end{eqnarray}
Since the torque is proportional to  $\mathbf{r} \times \hat{\mathbf{r}}_{m}$, for co-planar cases where $\mathbf{r}$ is always in the x-y plane, the tidal force just spins up (or down) the binary. The binary orientation should not change. However, if the binary is initially tilted, i.e. 
the binary axis is not parallel or anti-parallel to the z-axis, 
the binary orientation should change in general. 

The ratio of the binary angular momentum to the COM angular momentum is 
roughly given by 
\begin{eqnarray} \label{eq:angmomcomp}
\frac{L_{b}}{L_{m}} \sim \left(\frac{m}{M}\right)^{2/3} D^{-1/2},
\end{eqnarray}
where we have assumed equal-mass binaries. If we assume a typical central massive BH $\sim10^{6} M_{\odot}$ and a stellar mass binary, the ratio is of order $\sim 10^{-4} D^{-1/2}$. Even in very deep encounter cases (e.g. $D \sim 10^{-3}$), $L_{b}$ is much smaller than $L_{m}$. The flip of $\mathbf{L}_{b}$ does not affect $\mathbf{L}_{m}$ significantly, and this ensures the validity of the restricted parabolic approximation.
 
The effective spin is defined by
\begin{eqnarray} \label{eq:effspin}
\chi\textsubscript{eff} = \frac{1}{m}\paren{m_1\mathbf{S}_{1} + m_2 \mathbf{S}_{2}}\cdot \frac{\mathbf{L}_{b}}{|\mathbf{L}_{b}|},
\end{eqnarray}
where $\mathbf{S}_{1,2}$ are the dimensionless spins of the BHs in the binary, and they are bounded by $0 \le S_{1,2} < 1$.  The effective spin $-1< \chi\textsubscript{eff} < 1$ is a constant of motion, up to 
at least the 2nd post-Newtonian order \citep{blanchet14}, and it can be 
measured by GW observations. The distribution of effective spins is expected to shed light on the formation channels of BH mergers \citep{farr1, farr2, accgravdat, gerosa1}.

As we have mentioned, the dynamics of the tidal encounter does not directly depend on the masses of the binary members. Restricted three-body results can be simply rescaled in terms of their masses. However, we need to specify the mass ratio $m_1/m_2$ to evaluate the effective spin. Considering that the BH mergers detected by LIGO/Virgo to date are consistent with equal mass members, we assume $m_1=m_2$ when the effective spin $\chi\textsubscript{eff}$ is discussed. For simplicity, we also assume $S=|\mathbf{S}_{1}|=|\mathbf{S}_{2}|$.
Another simple case, $|\mathbf{S}_1|=S$ and $\mathbf{S}_2=0$, will be 
briefly considered in the discussion section.

If BH spins are initially parallel to $\mathbf{L}_{b}$ (this condition will be relaxed later), the effective spin
of a survivor indicates whether/how the binary orientation changes at the tidal encounter, and it is given by
\begin{eqnarray}
\chi\textsubscript{eff,out} = S ~\hat{\mathbf{L}\hspace{-2pt}}_{\; b, in} \cdot \hat{\mathbf{L}\hspace{-2pt}}_{\;b, out},
\end{eqnarray}
where $\mathbf{L}_{b, in,out}$ are the angular momenta of the pre/post-encounter binaries, and the hat indicates unit vectors. We have assumed that the BH spin vectors do not change at the tidal encounter, because the binary separation and the distances to the central massive BH are much larger than their event horizon scales.
General relativistic effects should be negligible especially in the short period of the tidal encounter. 

\section{Numerical Study} \label{sec:surv}
We numerically investigate the tidal encounters of BH binaries with a massive BH \citep{kobayashi1, addison2015, disrupGC}. To simplify our analysis, we limit the study to initially circular binaries.
The initial orientation of a  binary is determined by the unit vector $\hat{\mathbf{L}\hspace{-2pt}}_{b}=(\cos{\theta}, \sin{\theta}\cos{\varphi}, \sin{\theta}\sin{\varphi})$.  Assuming specific values of the penetration factor $D$ and the effective binary phase $\phi$, the binary is injected into a parabolic orbit at a distance $r_{m}=10 r_{t}$. 
 
\begin{figure} 
\includegraphics[width=\columnwidth]{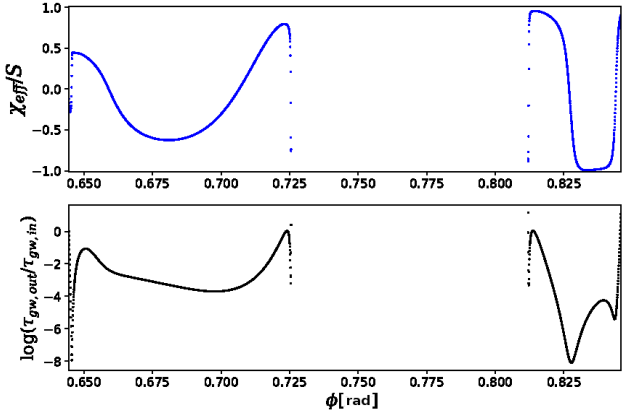}
\caption{Surviving binaries. Top panel: the post-encounter effective spin $\chi\textsubscript{eff}$ as a function of the effective binary phase $\phi$. Bottom panel: the post-encounter GW merger time $\tau_{gw,out}$ as a function of $\phi$.
$\chi\textsubscript{eff}$ and $\tau_{gw,out}$ are in units of the individual BH spin $S$ and the pre-encounter merger time $\tau_{gw,in}$, respectively. 
$D=0.5$, $\theta=0.5\pi$ and $\varphi=0.6\pi$ are assumed.}
\label{fig:difphis}
\end{figure}

The equation of motion (\ref{eq:eqmot2}) is integrated together with eq. (\ref{eq:eqmot3}) using a fourth order Runge-Kutta scheme. To ensure the accuracy of the dynamical evolution, at each instant the time-step width $\Delta t$ is chosen to be the smallest between the characteristic orbital time of the binary $t_{\textrm{bin}}$ and the free-fall time of the parabolic orbit $t_{\textrm{par}}$, multiplied by a normalization factor $h$. That is, $\Delta t = h \cdot \textrm{min}\{t_{\textrm{bin}}, t_{\textrm{par}} \}$  
\footnote{For the simulations described in this paper, we set the normalization factor to $h=10^{-3}$. We found that this was sufficient to adequately sample the binary evolution.}. 
If the system is coplanar, the binary orbit around its COM remains in the x-y plane at the tidal encounter. However, even a small inclination can lead to a significant change in the binary orientation. To illustrate this, we consider an almost coplanar case with the initial orientation $\theta=0.5 \pi$, $\varphi=0.6\pi$ and $D=0.5$. Note that prograde binaries have $\theta=0.5 \pi$ and $\varphi=0.5\pi$ ($\mathbf{L}_{b}$ is oriented in the z direction. See figure \ref{fig:angles}). In figure \ref{fig:difphis}, we plot the effective spin (the top panel) and GW merger time (the bottom panel) of the post-encounter binaries as functions of the effective binary phase $\phi$. Since we show only surviving cases, the gap between $\phi \sim 0.725$ and $\sim 0.81$ indicates that all binaries are disrupted in this range. We find that the binary orientation $\mathbf{L}_{b}$ flips to the almost opposite direction at the tidal encounter in the border regions, and the effective spins $\chi\textsubscript{eff}$ of the survivors can have large negative values if $S\sim 1$. Since disrupted binaries have $e > 1$, as we expect, the eccentricity 
and the semi-major axes of the survivors rapidly grow at the survivor boundaries. 
The wide binary separations (i.e. the longer lever arms) might help to induce a large torque in eq. \ref{eq:torque}, resulting in 
the negative effective spins at the boundaries. We find that survivors near the boundaries as well as inside the surviving region can have short GW merger times.  

\subsection{Survivors: the penetration factor dependence} \label{subsec:surv}
We first study how the properties of survivors depend on the penetration factor 
$D=r_p/r_t$, which is a key parameter to describe the tidal encounter dynamics. If the periastron $r_p$ is located well outside the tidal radius $r_t$, binaries should not be  affected by the tidal force of the massive BH at least during a single encounter. All binaries survive the tidal encounter if $D>2.1$. 
For smaller $D$, the surviving probability roughly linearly decreases $P_{sur} \propto D$ and it levels off at $P_{sur} \sim 10 \%$ around $D=0.1$ \citep{kobayashi1, brown1}.

Assuming that the binary orientation is isotropic and the binary phase is uniform, we evaluate the distributions of survivor properties for a given $D$. 
By taking into account the symmetry in the system, we assume that the binary orientations are uniformly distributed on the hemisphere defined by $0\le \theta \le \pi/2$ and $0\le \varphi < 2\pi$ \citep{brown1}. The effective binary phases $\phi$ are uniformly distributed between 0 and $\pi$ for each binary orientation \citep{kobayashi1}. 

\begin{figure} 
\includegraphics[width=\columnwidth ]{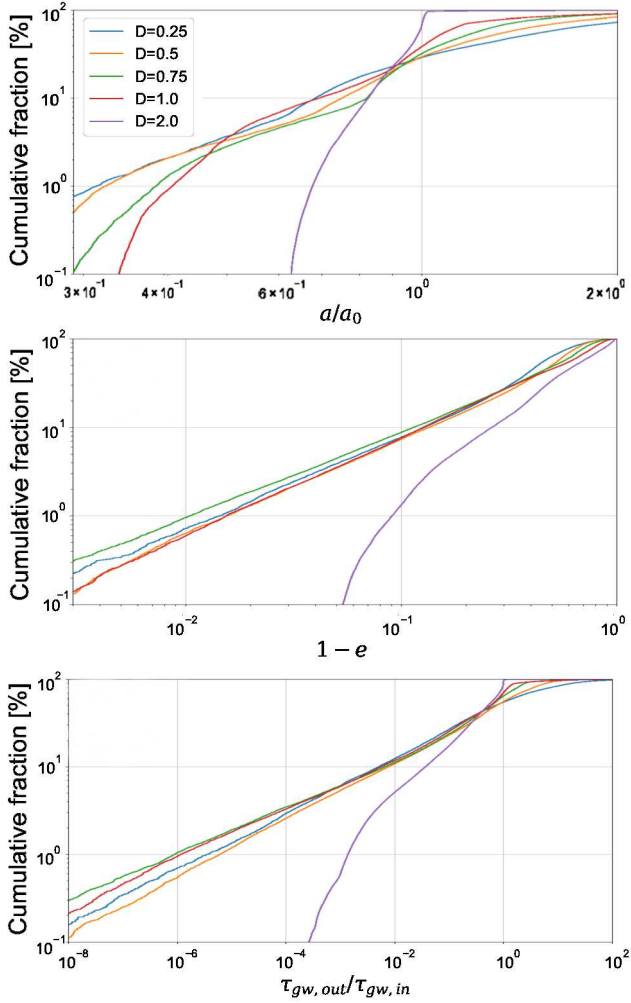}
\caption{Distributions of the semi-major axes $a$ (the top panel) and eccentricity differences $1-e$ (the middle panel) and GW merger times (the bottom) 
of the survivors. 
The semi-major axis $a$ and the GW merger time $t_{gw,out}$ are in units of the pre-encounter values of $a_0$ and $t_{gw,in}$.
The distributions are obtained from the Monte Carlo sampling with a fixed value of $D=0.25, 0.5, 0.75, 1.0$ or 2.0. }
\label{fig:orbital}
\end{figure}

Figure \ref{fig:orbital} shows the distributions of the semi-major axis (the top panel) and eccentricities (the middle panel) of survivors, which are obtained by randomly sampling 1000 binary orientations and  more than 200 binary phases. We have carried out the Monte Carlo sampling for $D=0.25, 0.5, 0.75, 1.0$ and 2.0. The distributions (especially the eccentricity distribution) are insensitive to $D$. Except the $D=2$ case, the distributions are similar to each other in each panel. For $\sim 3\%$ of the survivors, the semi-major axes are reduced by a factor of $>2$ from the pre-encounter separation $a_0$. The survivors are eccentric in general, and about $10\%$ of them have very high eccentricity $e>0.9$. 

The GW merger time greatly depends on the semi-major axis and eccentricity of the binary. We estimate the reduction factor of the merger time $\tau_{gw,out}/\tau_{gw,in}\equiv (a/a_0)^4 (1-e^2)^{7/2}$, which is the ratio of the survivor's merger time $\tau_{gw,out}$ to the pre-encounter one $\tau_{gw,in}$. 
The distributions of the reduction factors are shown in the bottom panel 
of figure \ref{fig:orbital}. The distributions are very similar to each other except the $D=2$ case. About $10\%$ ($1\%$) of the survivors have GW merger times shorter by a factor of $>100$ ($>10^5$) compared to the pre-encounter merger time.

\begin{figure} 
\includegraphics[width=\columnwidth,height=0.33\textheight]{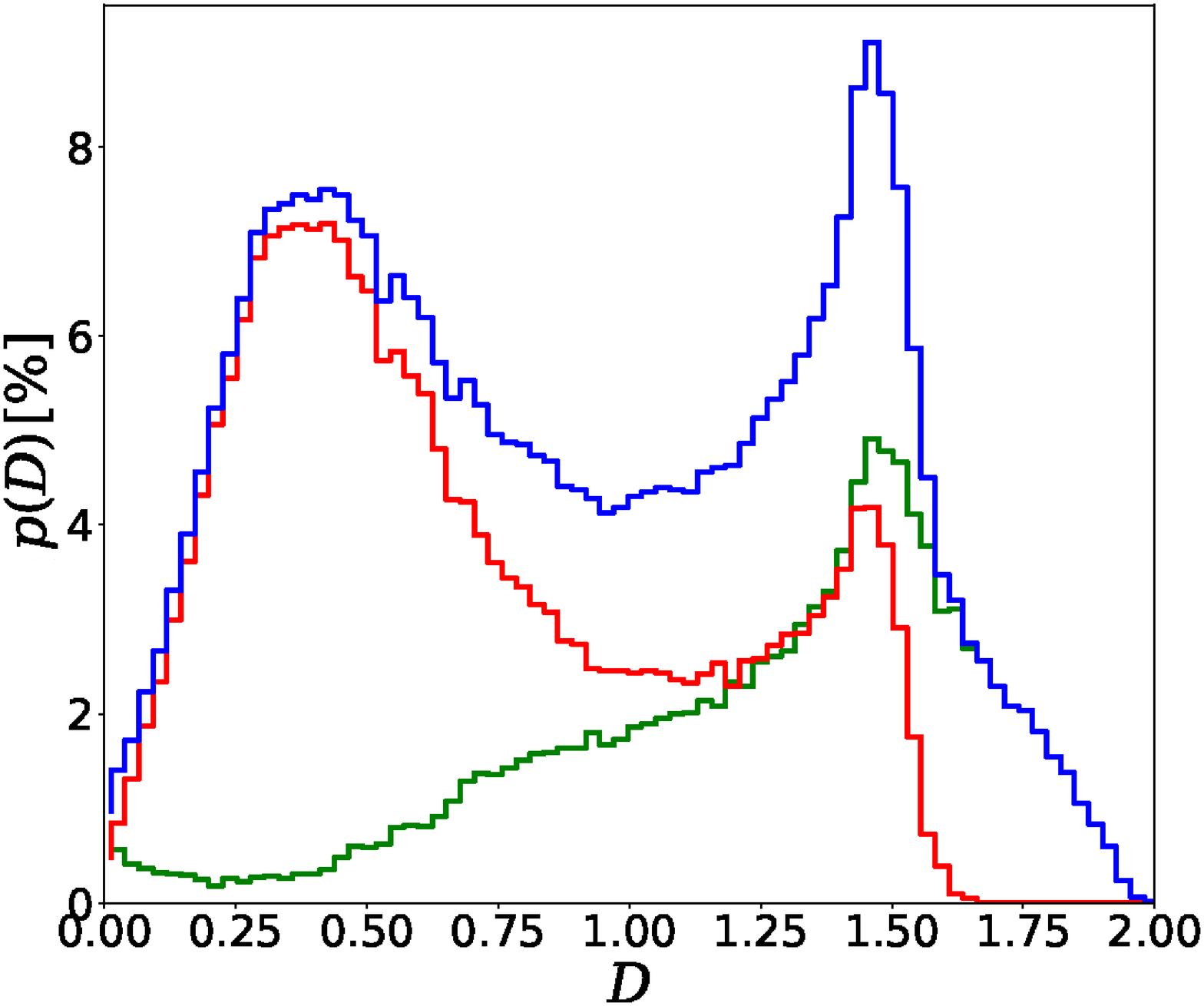}
\caption{Probability of survival with negative $\chi\textsubscript{eff}$ as a function of D.  The initial binary orientations are assumed to be isotropic (the blue line), 
prograde ($L_{b,z}>0$; the green line) or retrograde ($L_{b,z}<0$; the red line). 
}
\label{fig:negspin}
\end{figure}

The orientations of binaries can also change significantly at the tidal encounter. The blue line in figure \ref{fig:negspin} indicates the probability to get survivors with a negative effective spin as a function of $D$ (i.e. the probability that the binary survives the tidal encounter and the surviving binary has a negative effective spin when a binary with a random orientation and binary phase is injected with a given $D$). One finds that it is a bimodal distribution with a peak around $D=0.4$ and the other around $D=1.5$. Since the surviving probability is almost linear in $D$, the peaks indicate that a significant fraction ($\sim 40 \%$) of survivors have negative effective spins around $D=0.4$ (the fraction is about $10-15\%$ for $D=1-1.5$), and the fraction sharply drops for $D>1.5$. 

To investigate how the results depend on the initial binary orientation, we split the Monte Carlo sample into two groups, one for which the binaries are initially prograde ($L_{b,z} >0$) and one for which they are initially retrograde ($L_{b,z} <0$), where $L_{b,z}$ is the z-component of the pre-encounter angular momentum $\mathbf{L}_{b}$. The green and red lines
in figure \ref{fig:negspin} correspond to the prograde and retrograde cases, respectively. We have normalized their distributions as the sum of the two gives the total distribution, i.e. we have multiplied them by 1/2. We first notice that the peak around $D=0.4$ is due to the retrograde group (the red line).
Prograde binaries are known to be more vulnerable to the tidal disruption. Accordingly, the surviving probability for the prograde group rapidly decreases for deeper encounters $D<2.1$. Since the surviving probability is about a few $\%$ for the prograde group and about $40 \%$ for the retrograde group at $D=0.4$, the domination by the retrograde group is not surprising. However, since the surviving probability for the retrograde group is roughly linear in $D$ for $D<1.5$, it indicates that a good fraction ($\sim 40\%$) of retrograde binaries significantly change their orientations around $D=0.4$.

\subsection{The entire population of survivors}\label{subsec:entire}
BH binary populations in the Universe are still highly uncertain. 
The distribution of penetration factors $D$ is likely to be susceptible to the complicated galactic centre dynamics \citep{merrittrev, alexander, bradnick}. In general, one might expect comparable numbers of full and empty loss cone systems \citep{peretsalexander}. \cite{remi1} have recently shown that rare large scatterings can play a significant role, and the tidal encounter events which occur well inside the loss cone are almost as common as those with $D=1$ even in the empty loss cone regime. 
Here we assume two simple 
$D$ distributions: $P(D) \propto D^{\alpha}$ ($\alpha=0$ or 1) 
to illustrate our tidal encounter model.
These distributions correspond to situations close to the full loss cone regime. If $D\gg1$, the binary obviously survives the tidal encounter, and the properties of the binary do not change. 
We consider the range of $0<D<2.1$ to characterize the tidal encounter process. The choice of the threshold value
$D=2.1$ is motivated by our evaluation method of the BH merger rate.
In section 5, we will give a rough estimate of the BH merger rate 
due to the tidal encounter channel by using the tidal encounter rate of BH binaries with a massive BH. Since this tidal encounter rate will be inferred from the tidal disruption rate of stars or stellar binaries, rather than their tidal encounter rates, we have 
chosen $0<D<2.1$ (see section 5 for additional discussion).
Note that all binaries survive for $D>2.1$. 

As we have discussed in section \ref{subsec:surv}, the binary orientation $\{\theta, \varphi\}$ and the effective binary phase $\phi$ are assumed to be uniformly distributed. For each $D$ distribution 
($\alpha=0$ or 1), more than $4\times 10^{5}$ random realizations 
$\{D, \theta, \varphi, \phi\}$ are generated. We find that the surviving 
probability is 47 $\%$ for $\alpha=0$ and 54 $\%$ for $\alpha=1$.

\begin{figure*} 
\includegraphics[width=2\columnwidth,height=0.75\textheight]{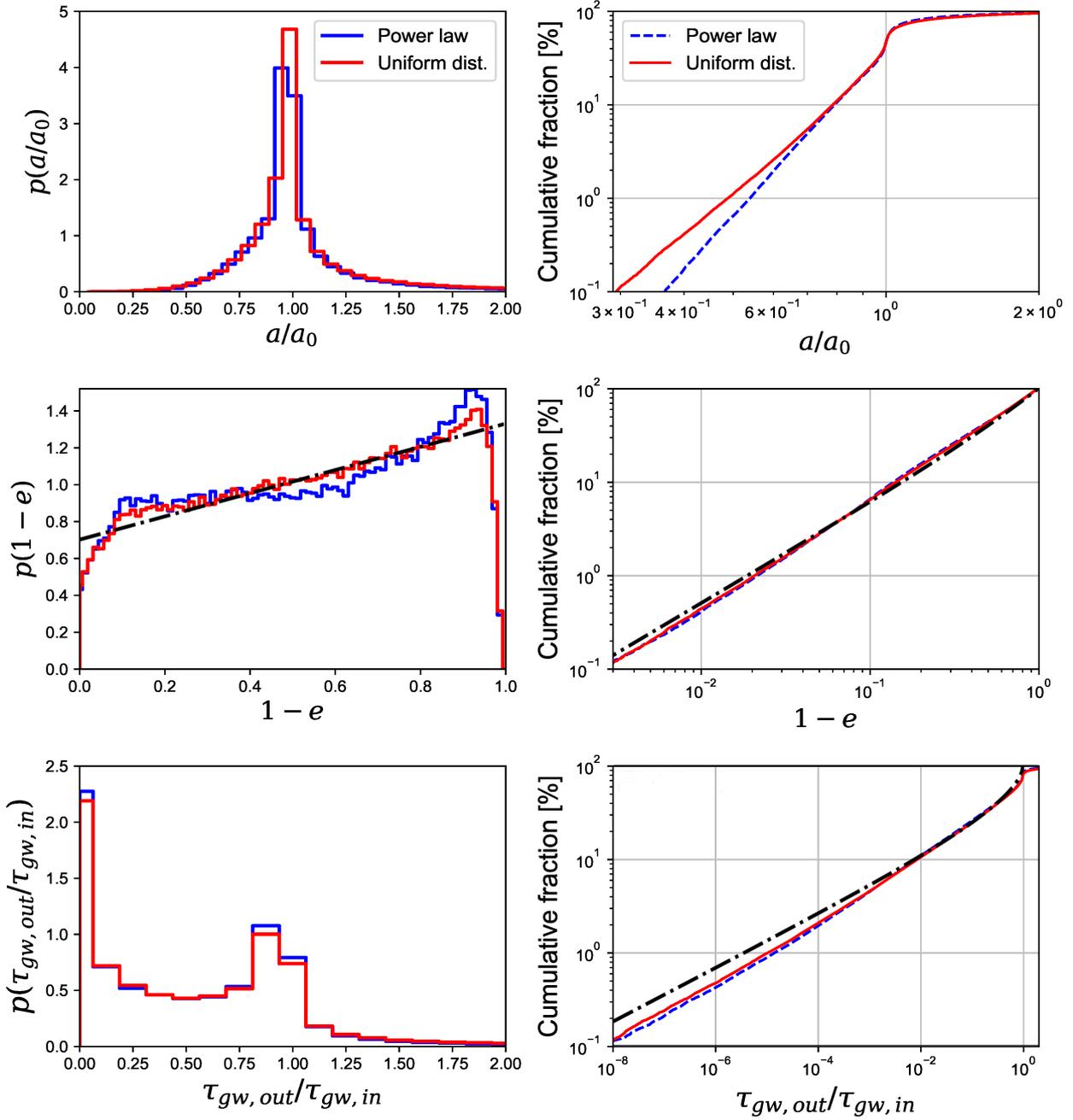}
\caption{Orbital parameters of survivors: 
$a$ (top panels),  $1-e$ (middle panels) and $t_{gw,out}/t_{gw,in}$ (bottom panels). The left panels indicate their probability distribution functions, and the right panels are for the cumulative distributions. 
The uniform $D$ distribution ($\alpha=0$) and the power-law distribution ($\alpha=1$) results are shown by the red solid and blue dashed lines, respectively. 
$a$ is in units of the initial separation $a_0$. The black dashed-dotted lines indicate a linear fit to the eccentricity distribution for the  $\alpha=0$ case (middle panels), 
and the analytic cumulative distribution of merger time reduction factors (bottom right panel).}
\label{fig:multiple}
\end{figure*}

Figure \ref{fig:multiple} shows the distributions of properties of
the survivors. Since the properties are rather insensitive to $D$ (as we can see in figure \ref{fig:orbital}) the two $D$ distribution models give similar results (the red solid line for $\alpha=0$ and the blue solid/dashed lines for $\alpha=1$). The distributions of the semi-major axes $a$ sharply peak at $a/a_0=1$ (the top left panel), and  $\sim 50\%$ of survivors have semi-major axes smaller than the initial value $a_0$. We find $a/a_0 < 0.5$ for about 1$\%$ of the cases.  The eccentricities of the survivors are more spread out (the middle left panel). About $50\%$ of the survivors have $e>0.5$, and several $\%$  have very high eccentricity $e>0.9$. 
These orbital changes significantly reduce the GW merger times of the binaries. The distributions of the merger time reduction factors are bimodal in the linear space (the bottom left panel). About $10\%$ of the surviving binaries have their merger times reduced by a factor of $10^{2}$ or more, and  about $1\%$ have very larger reduction factors of $>10^{5}$. 

It is primarily the growth in eccentricity, rather than the hardening of the binaries, which causes the GW-driven-merger timescale to shrink following tidal interactions. To illustrate this, we show that the $\tau_{gw,out}/\tau_{gw,in}$ distribution can be
reproduced from the eccentricity distribution. For the $\alpha=0$ case, the probability distribution function of the eccentricity can be fit with a linear function $P = (1+s)-2s e $ where $s\sim 0.31$. 
This linear function satisfies the normalization condition $\int_0^1Pde=1$. The probability distribution function of $(1-e)$ is shown as the black dashed-dotted line in the mid left panel of Fig \ref{fig:multiple}. We can see that the linear approximation describes the numerical results (the red solid line) reasonably well, except both ends $(1-e) \sim $ 0 or 1. Since the merger time reduction factor $\tau_{gw,out}/\tau_{gw,in}$ is proportional to 
$\xi \equiv (1-e^2)^{7/2}$, 
the distribution of $\xi$ would give that of the 
reduction factor if the binary hardening is negligible. 
Using the distribution function of eccentricity, 
we can evaluate the cumulative distribution function of $\xi$ as 
$P(>\xi)=\left(1+s\right) \left(1-\sqrt{1-\xi^{2/7}}\right)-s\xi^{2/7}$. 
The black dashed-dotted line in the bottom right panel shows 
the cumulative $\xi$ distribution where $\xi=\tau_{gw,out}/\tau_{gw,in}$ has been assumed to plot the function in the figure. The analytic function can describe the numerical reduction factor distribution (the red-solid 
line) reasonably well. It gives $1.3\%$ and $11\%$ 
for $\xi=10^{-5}$ and $10^{-2}$, respectively. 
These closely match the numerical results. 
The overestimate at low values $\tau_{gw,out}/\tau_{gw,in}\ll 1$
originates from the linear fit to the eccentricity distribution. Note that the linear distribution function $P(1-e)$ overestimates the numerical results as $e-1\to 0$. We can also analytically evaluate the (non-cumulative) distribution function of $\xi$ for $0\le \xi \le 1$, 
the function $P \propto \xi^{-5/7} (1-\xi^{2/7})^{-1/2}$ has a U-shape and 
peaks at $\xi=0$ and 1. In the bottom left panel, 
the high-side tail (i.e. the cases with $\tau_{gw,out}/\tau_{gw,in} > 1$) is due to survivors with $a>a_0$.

\cite{addison2015} study the properties of survivors, using full three-body calculations. Assuming a uniform $D$ distribution for $0.35 <D < 5$, they also have obtained the semi-major axes distribution very similar to ours (the top left panel of fig. \ref{fig:multiple}). In their sample, the majority of the surviving binaries are relatively unperturbed in eccentricity, but they have shown that a small fraction can have high eccentricity. 

\begin{figure*} 
\includegraphics[width=0.99\textwidth, height=0.32\textheight]{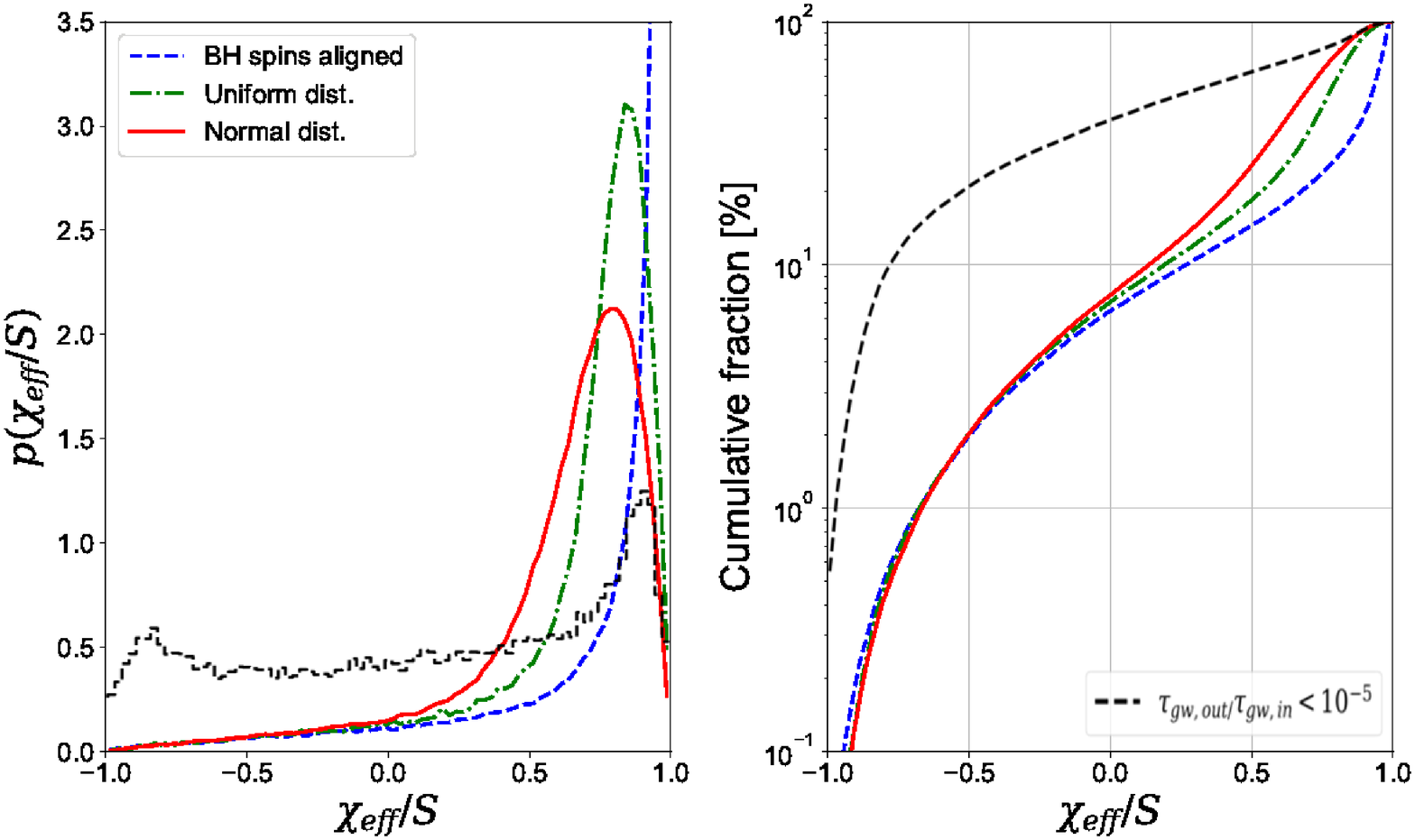}
\caption{Effective spin distributions of survivors (left panel) and their cumulative distributions (right panel). 
Initial BH spins are assumed to be  aligned with $\mathbf{L}_{b,in}$ (blue dashed line),  
uniformly distributed in the cone with opening angle $\pi/4$ around $\mathbf{L}_{b,in}$, 
or normally distributed with a standard deviation of $\pi/4$ around $\mathbf{L}_{b,in}$.
The distributions for survivors with $\tau_{gw,out}/\tau_{gw,in}<10^{-5}$ are also shown for the aligned spins case 
(black dashed line). The uniform $D$ distribution ($\alpha=0$) is assumed for all cases. The effective spin $\chi\textsubscript{eff}$ 
is in units of the BH individual spin $S$.}
\label{fig:withkick}
\end{figure*}

To estimate the effective spins of survivors, we have assumed that 
the spins of BHs in binaries are perfectly aligned with the 
pre-encounter binary angular momentum $\mathbf{L}_{b,in}$. 
We here consider additional cases to account for possible misalignment mechanisms (e.g. BH natal kicks). Although we still assume the same 
amplitude for the two BH spins $S=S_1=S_2$, the directions of the BH spins are now independent and random, uniformly distributed in the cone with opening angle of $\pi/4$ around $\mathbf{L}_{b,in}$, 
or normally distributed with a standard deviation of $\pi/4$ 
around $\mathbf{L}_{b,in}$, where $\mathbf{L}_{b,in}$ is the angular momentum of 
the pre-encounter binary. Figure \ref{fig:withkick} shows 
the effective spin distributions for the three BH spin models 
(aligned: the blue dashed line, uniformly distributed in the cone: the green dashed-dotted line, normally distributed in the cone: the red solid line). We find that the distributions are similar to each other for $\chi\textsubscript{eff}<0$. About $7\%$ of the survivors have negative effective spins. 

Although we have evaluated the effective spin distributions for the entire population 
of the survivors, only a fraction of them have short GW merger times, or more exactly speaking, significant reduction factors for the merger times. We have evaluated the effective spin distribution based on the aligned BH spin model for the survivors with reduction factors $\tau_{gw,out}/\tau_{gw,in} < 10^{-5}$. The resultant distribution (the black dashed-line)
is much flatter (see the left panel), and 39$\%$ of the population has negative effective spins. We also find that 19 $\%$ of survivors with $\tau_{gw,out}/\tau_{gw,in} < 10^{-2}$ have negative effective spins.

\section{Constraints from effective spin measurements} \label{sec:comp}

The effective spins of the BH mergers observed by LIGO/Virgo so far are clustered around $\chi\textsubscript{eff} \sim 0$, they are consistent with low effective spins 
within $-0.42 < \chi\textsubscript{eff} < 0.41$ at the 90$\%$ credible level 
(e.g \cite{bel17}, see also \cite{abbott16, ligo3, lv1, det5}). 
The positive effective spin of GW151226 $\chi\textsubscript{eff}=0.21^{+0.20}_{-0.10}$ 
indicates that at least one of the BHs in the binary has been spinning before the merger. 
In the other events, $\chi\textsubscript{eff}$ is consistent with zero within errors. The small values of the effective spins $\chi\textsubscript{eff}$
can result from  small BH spins $S$. If the intrinsic spins are almost zero for most BHs in binaries,  the current 
and future effective spin measurements would not give strong
constraints on the formation models of BH binaries.  However, 
if the intrinsic spins are large for a significant fraction of BHs, 
effective spin measurements  could reveal their origins.

BH spins in isolated field binaries are expected to be preferentially aligned with the orbital angular momentum.  Although natal kicks (e.g. anisotropic SN explosions or neutrino emission) can induce misalignment \citep{lignatkicks}, significant kicks would disrupt the binaries.  
It should be difficult to produce mergers with large negative  $\chi\textsubscript{eff}$. A non-vanishing fraction of mergers should have large positive $\chi\textsubscript{eff}$  if the intrinsic spin $S$ is large \citep{hoto17}.

BHs in dynamically formed binaries in dense stellar environments are expected to have spins distributed isotropically. The $\chi\textsubscript{eff}$ distribution is expected to be symmetric about zero, and it can be extended to high negative (or positive) $\chi\textsubscript{eff}$ 
if the intrinsic spin $S$ is large. 
Considering GW151226 with
$\chi\textsubscript{eff}>0$ and no definitive systems with $\chi\textsubscript{eff}<0$, the current sample is very weakly asymmetric. About $10$ additional detections are expected to be sufficient to distinguish between a pure aligned or isotropic population \citep{farr2}. 
 
In our tidal encounter model, a significant fraction of mergers can have negative 
effective spins $\chi\textsubscript{eff}$ especially if we consider the 
binaries with large reduction factors of the merger time. The  $\chi\textsubscript{eff}$ distribution is slightly asymmetric, but rather flat with minor enhancement at the high and low ends $\chi\textsubscript{eff}\sim \pm S$.

\section{Discussion and conclusions} \label{sec:conc}
We have studied how the tidal encounter with a massive BH affects the properties of BH-BH binaries (e.g. GW merger times and effective spins). Since we have treated binary members as point particles, our results are also applicable to the study of other types of compact stellar binaries (BHs, neutron stars and white dwarfs).

Binaries can survive the tidal encounter even in the deep limit $D\ll 1$. Although deep encounter survivors are counter-intuitive, binaries are actually disrupted, and the binary members separate when they deeply penetrate the tidal sphere of the massive BH. However, they approach each other after the periastron passage and a small fraction of them ($12\%$ for $D\ll 1$) can form binaries again even in the deep penetration cases \citep{kobayashi1, brown1}.

Assuming the simple $D$ distribution models (i.e.  an uniform or linear distribution for $0<D<2.1$), we have shown that about $50\%$ of injected binaries can survive the tidal encounter, and the GW merger times of the survivors can be shorter by many orders of magnitudes than those of pre-encounter binaries. About $10\%$ ($1\%$) of the survivors have GW merger times shorter by a factor of $>100$ ($>10^5$) than those of the pre-encounter binaries.  Assuming that BH spins are aligned with the binary angular momentum before the tidal encounter, we have shown that survivors can have negative effective spins. 
This is because the tidal force rotates the orientations of binaries, and the orientation flips 
to the opposite direction in some cases. Since BH spins are only weakly constrained by observations (and observations are consistent with equal mass mergers), for simplicity, we have assumed 
equal masses and 
equal BH spin magnitudes for binary members, $m_1=m_2$ and 
$S=|\mathbf{S}_{1}|=|\mathbf{S}_{2}|$. As the BH spins are constant in our tidal encounter model, it is straightforward to examine other models. For example, in another equal mass case of 
$|\mathbf{S}_1|=S,\ \mathbf{S}_{2}=0$ (e.g \cite{doron, matias}),
the effective spins $|\chi_{\textrm{eff}}|$
of survivors are maximally $S/2$, rather than $S$.  
Since we have assumed that the directions of BH spins are 
independent and random when the effective spins are evaluated for the uniform and normal distributions of BH spins, the effective spin
distributions should be identical to those shown in 
figure \ref{fig:withkick} if the x-axis is rescaled
(i.e. $\chi_{\textrm{eff}}/S$ should take a value 
between -0.5 and 0.5, all the distribution peak around 
$\chi_{\textrm{eff}}/S\sim 0.5$).

Although we have mainly discussed the tidal encounter survivors, a large fraction of binaries break up at the encounter. In such cases, one of the binary members should be ejected as a hyper-velocity BH and the other captured in a highly eccentric orbit around the massive BH. This is one of possible channels to produce 
extreme-mass-ratio inspirals \citep{miller05}, which are promising GW sources for the LISA mission \citep{lisav}. The tidal capture of BH binaries also has been discussed \citep{chen18}.

It is not trivial to estimate how frequently BH binaries merge due to the tidal encounter mechanism. Several processes are involved in the estimate, most of which are not well constrained by current observations \citep{miller05}. 
We here make a rough estimate of the merger rate due to the tidal encounter channel as
\begin{eqnarray}
\mathcal{R} \approx n_{g} \cdot \mathcal{N} \cdot P,
\end{eqnarray}
where $n_{g}$ is the number density of galaxies, $\mathcal{N}$ is the tidal encounter rate of BH binaries with a massive BH (events per yr per galaxy),  
$P$ is the fraction of tidal encounters that produce survivors with $\tau_{gw}<10^{10}$ years. We assume that the first galaxies formed about $10^{10}$ years ago, and they have had sufficient time to host and grow massive BHs. This is consistent with recent observations, which indicate quasars are known to exist when the Universe was less than a billion years old \citep{banadosquasar}. 
Since survivors merge many years after the tidal encounters, this estimate implicitly assumes that the merger rate reaches a steady state. 

The fraction $P$ depends on the semi-major axis distribution of the pre-encounter circular binaries. As galactic centres are collisional environments, 
wide binaries can be disrupted by encounters with other objects. 
Equalizing the binding energy $Gm_1m_2/2a$ with the kinetic energy of an intruder $m_\ast\sigma^2/2$, we obtain $a=Gm_1m_2/m_\ast\sigma^2 \sim 140$ au for $m_1=m_2=30M_\odot$, $m_\ast = 1M_\odot$ and 
the Milky-Way velocity dispersion $\sigma \sim 75$ km/s \citep{BHMsigma}. We set the maximum semi-major axis at this value. The minimum semi-major axis is set at $a_{0} = 0.2$ au for which binaries with $m_1=m_2=30M_\odot$ do not merge within $10^{10}$ yrs if they are not disturbed by the tidal encounter or other mechanisms.  
These binaries emit weak GW at low frequencies
$f_{gw} < 5.5\times 10^{-6} (m/60M_\odot)^{1/2}(a/0.2 ~\mbox{au})^{-3/2}$ Hz. Assuming a uniform $a_0$ distribution in logarithmic space, and 
using the $a/a_0$ and $e$ distribution for $\alpha=0$ obtained in section \ref{subsec:entire}, we evaluate the merger time $\tau_{gw}$ distribution of survivors (the $\alpha=1$ case  also gives a very similar distribution). We find that $\sim 50\%$ of BH binaries survive the tidal encounter and $\sim 6\%$ of the survivors have merger times of less than $10^{10}$ years, yielding $P\sim 3\times 10^{-2}$. 
We also have evaluated the factor $P$ by assuming that initial binaries are eccentric (a uniform or thermal distributions of initial eccentricity). Our preliminary results indicate that the fraction $P$ is very similar (Fernandez et al. in preparation).

Although the tidal encounter rate is highly uncertain, we adopt $\mathcal{N}=10^{-6}$ /yr/galaxy as a fiducial value. Stars are tidally disrupted by a massive BH with a rate of 
$10^{-5}-10^{-4}$/yr/galaxy \citep{komossa}. In the 
Milky Way, hypervelocity stars and the S-star cluster imply a similar rate of $10^{-5}-10^{-3}$/yr/galaxy for the disruption of stellar binaries \citep{bromley}. Simulations of galactic dynamics indicate that a density cusp forms around MBHs, where the concentration of high-mass objects increases. Population synthesis predictions also suggests that the fraction of BHs and NSs present in these regions is enhanced with respect to the field. In particular, following the simple formalism presented in \cite{bence2019}, it can be shown that the fractions of BHs in the entire Milky Way and the nuclear star cluster are $\sim 0.13\%$ and $\sim 0.23\%$ respectively \citep{licquia}. In addition, it is expected that dynamical friction will drive BHs formed in the outer regions towards the centre, further increasing their number \citep{lockman, petrovich}. Recent numerical estimates have shown that this effect can increase their number by a up to factor of several \citep{bence2019}. These results are supported by recent observations of quiescent X-ray binaries in the Milky Way galactic centre, indicative of a large population of BHs and BH binaries in the galactic centre \citep{xrbnat}. Hence the tidal encounter rate of compact binaries would be smaller than that inferred from hypervelocity star observations for stellar binaries by a factor of $\sim 10^2$.

In the early Universe, the number density of galaxies was higher, but most of these galaxies were relatively small and faint, with masses similar to those of the satellite galaxies surrounding the Milky Way (e.g.
\citep{conselice}). Assuming the galaxy number density $n_{g} \sim 0.02\ \textrm{Mpc}^{-3}$ \citep{conselice2005, mergerskozai}, we obtain
\begin{eqnarray}
\mathcal{R} \approx 0.6 ~\textrm{Gpc}^{-3} \textrm{yr}^{-1}.
\end{eqnarray}
This is much smaller than the BH merger rates inferred by GW observations $9.7-101 ~\textrm{Gpc}^{-3} \textrm{yr}^{-1}$ \citep{det6}. The tidal encounter mechanism is unlikely to be the dominant formation channel of BH mergers. However, the current and near-future GW observatories are expected to detect an enormous number of BH mergers. A small fraction of them might have experienced the tidal encounter with a massive BH.  The merger times of hard binaries (originally $\tau_{gw} \ll 10^{10}$ years) can be further shortened by this mechanism. If binaries merge in the vicinity of massive BHs, GW lensing echoes might be produced \citep{kocsisecho}. 

Our simulations show that the GW merger time can be 
reduced even in shallow tidal encounter cases with $D>2.1$.  However, the reduction effect becomes insignificant quickly for a larger value of $D$ (e.g. for $D=2.5$, GW merger times are reduced by a factor of $>10$ ($>50$) only in $10\%$ ($1\%$) of cases, 
for $D=3$, more than $99\%$ of post-encounter binaries have $\tau_{\textrm{gw,out}}/\tau_{\textrm{gw,in}}\gsim 0.8$). If we consider the range $0<D<3$, we find that the survivor fraction is $\sim 66\%$. The main result in section \ref{sec:surv} is modified by a factor of two: $\sim 5\%$ ($\sim 0.5\%$) of survivors have the merger times reduced by a factor $>100$ ($>10^{5}$). Although this might affect the merger rate estimated above, the uncertainty in the tidal encounter rate is much larger.

We now consider whether it is possible to observe tidal encounter events by using GW detectors.  Such observations might provide constraints on 
the very uncertain tidal encounter rate.
The primary GW signals from  encounters have been discussed as extreme-mass-ratio bursts (EMRBs) \citep{turner1977, rubbo, bg429}. They are produced when a binary, which can be treated as a point particle at the lowest order of approximation, passes by a massive BH. 
The GW signal will thus have a burst-like behavior, roughly characterized by an amplitude $h_B \sim G^2Mm/c^4r_p d$ and 
a duration $\Delta t \sim 1/f_{B} \sim \sqrt{r_p^3/GM}$ (e.g. \cite{KLPM04}).  Expressing the periastron distance $r_p$ in terms of the GW frequency $f_B$, we get 
\begin{eqnarray}
h_B  \sim 10^{-21}
\left( \frac{M}{4\times10^{6}M_{\odot}}\right)^{2/3}
\left(\frac{m}{30 M_{\odot}} \right)
\left( \frac{f_{B}}{10^{-3}\mbox{Hz}}\right)^{2/3}
\left( \frac{d}{10^2 \textrm{Mpc}}\right)^{-1}.
\end{eqnarray}
LISA would be able to detect EMRBs from massive BHs  out to $\sim 100$ Mpc \citep{bg433, sensecurves}.  

If the periastron distance and tidal separation are comparable (i.e. tidal encounters, $r_p \sim r_t$), the pre-encounter circular binary emits GWs at a frequency $f_b$ similar to the EMRB frequency $f_B$, but with an amplitude smaller by a factor of $\sim (M/m)^{2/3}$. 
Since the signal-to-noise ratio can be enhanced by integrating the periodic signal, the effective amplitude of the GWs from the binary would be $\sim \sqrt{N_c} (M/m)^{-2/3} h_B$  
where $N_c=f_{b} \Delta T_{obs}$ is the number of cycles radiated during an observational period $\Delta T_{obs}$.   In order to make the effective amplitude comparable to that of the EMRB, we need a very long observational run with
$\Delta T_{obs} \sim f_{b}^{-1} (M/m)^{4/3}\sim 200$ yrs for $M=4\times 10^6 M_\odot$, $m=30 M_\odot$ and $f_{b}=10^{-3}$ Hz. For a more realistic observational period $\Delta T_{obs} =1-5$ yrs, the effective amplitude would be smaller by 
one order of magnitude than the EMRB's amplitude. 
Therefore, EMRBs would be the dominant GW signals in tidal encounter
events, and they could indicate how frequently compact objects pass
by massive BHs. However, it could be difficult to distinguish binary encounter events from single object encounter events. 

If an equal-mass circular binary with $m=30 M_\odot$ is emitting GWs at $f_b=10^{-3}$ Hz, its GW merger time is $\sim 3\times 10^4$ yrs.
This binary can merge within the age of the Universe without the aid of tidal effects. 
It takes about $\sqrt{r_h^3/GM}\sim 8\times 10^3 (M/4\times 10^6 M_\odot)^{-1/2}(r_h/1\mbox{pc})^{3/2}$ yrs for the binary to travel from the radius $r_h$ of the BH sphere-of-influence to the centre. The binary can reach the tidal radius of the central BH before merging. 

If solar-type stars pass by a massive BH with periastron distances
comparable to their tidal radii $\sim (M/m_\ast)^{1/3} r_\ast$
where $m_\ast \sim M_\odot$ and $r_\ast \sim R_\odot$ are the mass and radius of the star, these produce 
EMRBs with $f_B \sim \sqrt{GM_\odot/R_\odot^3} \sim 10^{-3}$ Hz. 
The GW amplitude would be smaller by a factor of $\sim m/m_\ast$, 
and the EMRBs might be associated with electromagnetic counterparts, tidal compression flares \citep{KLPM04} and subsequent tidal disruption flares \citep{komossa}. Although very massive BHs $M\gsim 10^8M_\odot$
are known to swallow solar-type stars without tidal disruption
(i.e. their event horizon scale is larger than the tidal radius),
we can give similar discussion for binaries.  The 
ratio between the tidal radius of a binary and the event horizon scale of a massive BH
can be given in a simple form as 
\begin{eqnarray}
r_t/r_g&=& (c^2/2)\paren{\pi GMf_{b}}^{-2/3},\\
&\sim & 3.2 ~
\paren{\frac{M}{4\times10^6M_\odot}}^{-2/3}
\paren{\frac{f_{b}}{10^{-3} \mbox{Hz}}}^{-2/3} 
\end{eqnarray}
where a circular binary has been assumed, and this ratio does not depend on the binary mass. 
Binaries emitting GWs at $f_b\gsim 6\times 10^{-3}$ Hz would be 
swallowed by $4\times 10^6M_\odot$ BHs without tidal disruption.

\section*{Acknowledgements}
We thank the anonymous referee for valuable suggestions and Bence Kocsis for useful discussions.  This research was supported by STFC grants and a LJMU scholarship.


\bsp	
\label{lastpage}
\end{document}